\documentclass{article}

\usepackage{arxiv}

\usepackage[utf8]{inputenc} 
\usepackage[T1]{fontenc}    
\usepackage{url}            
\usepackage{booktabs}       
\usepackage{amsfonts}       
\usepackage{nicefrac}       
\usepackage{microtype}      
\usepackage[hidelinks]{hyperref}
\usepackage{cleveref}       
\usepackage{lipsum}         
\usepackage{graphicx}
\usepackage[square,numbers]{natbib}
\usepackage{doi}

\usepackage{moreverb,url}
\usepackage{comment}
\usepackage{tabularx}
\usepackage{soul}

\usepackage{caption}
\usepackage{subcaption}

\newcommand{\imageWidthSmall}{0.30}

\title{Using Text-to-Image Generation for Architectural Design Ideation}

\renewcommand{\shorttitle}{Using Text-to-Image Generation for Architectural Design Ideation}

\author{ \href{https://orcid.org/0000-0000-0000-0000}{\includegraphics[scale=0.06]{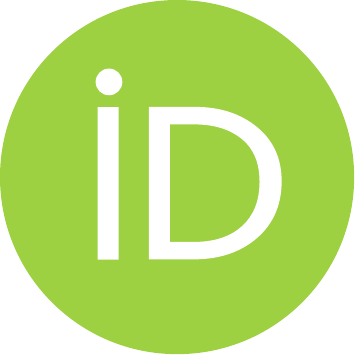}\hspace{1mm}Ville Paananen} \\
	Center for Ubiquitous Computing\\
	University of Oulu\\
	Oulu, Finland \\
	\texttt{ville.paananen@oulu.fi} \\
	\And
	\href{https://orcid.org/0000-0000-0000-0000}{\includegraphics[scale=0.06]{orcid.pdf}\hspace{1mm}Jonas Oppenlaender} \\
	Faculty of Information Technology\\
	University of Jyväskylä\\
	Jyväskylä, Finland \\
	\texttt{jonas.x1.oppenlander@jyu.fi} \\
	\And
	\href{https://orcid.org/0000-0000-0000-0000}{\includegraphics[scale=0.06]{orcid.pdf}\hspace{1mm}Aku Visuri} \\
	Center for Ubiquitous Computing\\
	University of Oulu\\
	Oulu, Finland \\
	\texttt{aku.visuri@oulu.fi} \\
}

\hypersetup{
pdftitle={Using Text-to-Image Generation for Architectural Design Ideation},
pdfauthor={Ville Paananen, Jonas Oppenlaender, Aku Visuri},
pdfkeywords={Architecture, text-to-image generation, generative AI, design creativity},
}

\begin{document}
\maketitle

\begin{abstract}
The recent progress of text-to-image generation has been recognized in architectural design. Our study is the first to investigate the potential of text-to-image generators in supporting creativity during the early stages of the architectural design process. 
We conducted a laboratory study with 17 architecture students, who developed a concept for a culture center using three popular text-to-image generators: Midjourney, Stable Diffusion, and DALL-E. Through standardized questionnaires and group interviews, we found that image generation could be a meaningful part of the design process when design constraints 
are carefully considered. Generative tools support serendipitous discovery of ideas and an imaginative mindset, enriching the design process. We identified several challenges of image generators and provided considerations for software development and educators to support creativity and emphasize designers' imaginative mindset. By understanding the limitations and potential of text-to-image generators, architects and designers can leverage this technology in their design process and education, facilitating innovation and effective communication of concepts.
\end{abstract}



\section{Introduction}

In the field of architecture, effective ideation hinges on the ability to represent ideas. Traditionally, drawings, photographs, and other visual media have been used to stimulate ideation and communicate design concepts. However, recent advances in generative artificial intelligence (AI) have made it possible to generate detailed and realistic representations of architectural concepts, using prompting in natural language as a general-purpose interface~\cite{nichol2022glide,pmlr-v139-ramesh21a,stablediffusion,imagen,2102.07350.pdf}.
The use of generative AI and procedural design is not new to architecture, and their use date back to the 1970s~\cite{gero}.
However, prompt-based generation marks a paradigm shift that could affect the architectural design process.
Text-to-image generation tools \cite{stablediffusion,dalle2} are one such example of generative AI. Text-to-image generation tools can allow for a quick conceptualization of ideas with natural language during the idea generation process. Thus, these tools have the potential to transform the way architects and designers develop and communicate their ideas.

In this paper, we study how different text-to-image generators can support creativity during the ``fuzzy front end''~\cite{fuzzyfrontend} of new concept development in the early stages of the architectural design process.
In particular, we investigate:%
\begin{enumerate}%
    \item How can text-to-image generators support creativity and ideation during the early stages of architectural design?
    \item How effective are out-of-the-box text-to-image generators in the context of architectural design, and what future considerations could developers take into account?
    \item What are the typical challenges of text-to-image generator use and text prompting for novel users?%
\end{enumerate}

In a laboratory study, 17~participants developed a concept for a culture center using three popular text-to-image generators; Midjourney~\cite{midjourney}, Stable Diffusion~\cite{stablediffusion}, and DALL-E~\cite{dalle2}. Through standardized questionnaires on creativity support tools and group interviews, we learned that image generation could be a meaningful part of the design process when design constraints and imaginative ideation are carefully considered. Generative tools support the serendipitous discovery of ideas and an imaginative mindset, which can enrich the design process. Through our study, we highlight several challenges of image generators and provide considerations for software development and educators to support creativity and emphasize designers' imaginative mindset.


\section{Related work}

To contextualize the use of text-to-image generators in architectural design, we first describe the different ways creativity has been approached in the architectural design process. Then, we present current literature on how creativity-supporting generative tools have been applied in the architectural design process.

\subsection{Creativity in the architectural design process}
Architecture is highly relevant to creativity as a field concerned with solving problems in contextual and effective ways~\cite{taneriHowLearnBe2021}
The prior literature has developed multiple notions for understanding what aspects affect creativity and how it can be supported in architectural design. For instance, Sarkar and Chakrabarti formulated creativity as the function of novelty and usefulness, allowing for the assessment of different design outcomes~\cite{sarkarAssessingDesignCreativity2011}.
This two-factor definition of creativity into
    novelty (or its related synonyms, such as originality, unusualness, or uniqueness)
    and usefulness (or its related synonyms effectiveness, fit, or appropriateness)
    -- has become popular in scholarly literature as a way to operationalize and measure the concept of creativity~\cite{StandardDefinition}.
However, creativity in architecture is not only related to producing a final outcome that is novel and useful but it is also related to the application of one's creative skills during the creative design process.
Casakin et al. found that creative thinking skills are more related to verbal skills rather than figural skills~\cite{casakinCreativeThinkingPredictor2010}.
Consequently, the authors proposed that the creative skills in architecture are also generalizable to other problem-solving areas in life. Additionally, Baghai Daemei \& Safari found that students' experience of design processes is a critical aspect of creativity~\cite{baghaeidaemeiFactorsAffectingCreativity2018}.

Creativity has also been operationalized in architecture using various tools. Park et al. studied how text stimuli can improve a person's imagination in nonlinear architectural design tasks~\cite{parkArchitectureImaginaryText2022}.
    Using text snippets from Italo Calvino's \textit{Imaginary Cities}, the partaking architecture students produced imaginative concepts. The findings show that text stimuli can support nonlinear creativity and move the design focus from the outcome to the process. As such, the multimedia approach requires designers' imagination.
In order to assess artifact creativity, Demirkan \& Afacan used exploratory and confirmatory factor analysis to develop a descriptive set of 41 design creativity words~\cite{demirkanAssessingCreativityDesign2012}.
    The design elements were then grouped into three factors: ``Artifact creativity,'' ``Design elements,'' and ``Assembly of creativity elements.'' 
Kowaltowski et al. interviewed architecture educators about stimulating creativity in architectural design~\cite{kowaltowskiMethodsThatMay2010}. The resulting 
creativity support methods show that generative methods are relevant for producing many unfamiliar ideas, but it also requires more support from teachers. Similarly, Mose Biskjaer et al. developed an analytical framework for understanding how creativity methods are used in design processes~\cite{mosebiskjaerUnderstandingCreativityMethods2017}.
The authors suggest that the design process itself can be designed through concrete, conceptual, and design spaces aspects. As such, reformulating the design task can help to bring forward more creative solutions. 

\subsection{Text-to-image generation in architectural design}

Text-guided diffusion models~\cite{nichol2022glide,pmlr-v139-ramesh21a,stablediffusion,imagen} have become a popular means of synthesizing novel images from input prompts written in natural language, and generative AI is increasingly being employed in academia and in the industry.
Generative AI in architectural design has been explored in two surveys. Through reviewing machine learning research trends in architecture, Ozerol \& Arslan Selçuk found that generative AI is rising in popularity~\cite{ozerolMachineLearningDiscipline2022}.
However, machine learning was more often applied in 3D generative methods than in 2D. This suggests that the fast development of text-to-image generative methods has not yet reached the architectural research community.
In a survey of generative systems in architectural, engineering, and construction research between 2009 and 2019, BuHamdan et al. found that many generative methods are focused on architectural, structural engineering, and urban design disciplines~\cite{buhamdanGenerativeSystemsArchitecture2021}. However, the most popular architectural use cases (facade design, form generation, layout generation) represent more geometric processes, and the role of more conceptual creativity in generative systems is still left to be unexplored.

Text-to-image generation systems provide an easy-to-use interface due to the ability to respond to natural language prompts. However, the creativity of text-to-image generation currently still hinges on the skill of its users~\cite{TTI-creativity}.
To control the output, users have to resort to special keywords in the prompts to produce images in a certain style or quality~\cite{TTI-taxonomy}.
Longer prompts also typically produce images of higher quality~\cite{2303.04587.pdf}.
While text-to-image generation tools can be intuitive, their application in the context of architecture remains yet to be explored.

Seneviratne et al. used a systematic grammar to explore the robustness of a text-to-image generator in the context of the built environment~\cite{seneviratneDALLEURBANCapturingUrban2022}.
The study found that the image generator was broadly applicable in the context of architecture.
However, architectural semantics contain ambiguities~\cite{bolojanLanguageAllWe2022},
and the real-world benefit of text-guided image generation remains to be explored.
In this paper, we specifically focus on how text-to-image generation tools can support human divergent creativity during the early-stage concept design process.



\section{Method}
We designed a laboratory study in which architecture students engaged with text-to-image generators in a short architectural design task.
The study was tested in a formative pilot study in which two authors and three colleagues used text-to-image generators to create their dream home. The formative pilot informed the design of the main study design, as follows.

\subsection{Study design and procedure}
We conducted three sessions (henceforth \textbf{S1}, \textbf{S2}, and \textbf{S3}) where 5--6 participants each individually worked on the same task. Participants were tasked to design a concept for a culture center. The site is a small island with a small observatory and interconnecting bridges to the nearby downtown area.
Participants were tasked to brainstorm visual concepts supporting the overall design task. More specifically, participants were asked to produce visual representations of their concept, including 1) a floorplan, 2) an interior perspective visualization, and 3) a facade material sample. Participants could use pen and paper to support their ideation, but the generated images could not be edited digitally.

The study procedure was as follows. Participants were first asked to provide informed consent. Participants were then given an introduction to text-to-image generators.
The text-to-image user interfaces were not modified in any way.
Participants were introduced to the tools with a short presentation on how to use the text-to-image generators for architecture and some unrelated example images produced by each of the three tools. Participants were then given a short interactive tutorial task that revolved around generating and iterating an increasingly complicated image of a pineapple. Only basic functionalities -- text prompts and generating variations or upscaled versions of generated images  -- of the tools were allowed during the session to ensure comparability of the three tools. Advanced features that participants were not allowed to use during the study include features like inpainting or using generated images as the basis for future generations. These features were omitted since not all of the three tools contain all advanced features. 
Participants were then presented with a design brief and began working on the task.

Participants had between 1h 15min and 1h 25min to work on their task. The duration varied according to how long the initial stages of the study lasted and at what point the participants felt they had completed the task. In each session, all three tools were used, with 1--2 participants each using one of the three tools. Each participant had their own laptop (either personal or borrowed) to work with. While they worked individually towards their own solution, participants could talk and discuss freely amongst each other, as well as take breaks when needed. This decision was taken to emulate a collaborative work environment in an organization and to support the participants' individual creative needs. 
One researcher was present at all times to answer questions.
A second researcher acted as an observer, taking notes of any interesting discussions and observations during the design session.
The sessions were recorded using a conference microphone that could capture the audio in the whole meeting room. Once participants finished working on the design task, each participant presented their work to the group, and to further motivate the participants, the participants voted for the best work using ranked choice and ranking their own design lowest. Participants were compensated with a 15 EUR 
gift card and the winner of each session was awarded an additional 15 EUR gift card.%
\subsection{Data collection}%
At the end of the session, we administered the Creativity Support Index (CSI)~\cite{cherryQuantifyingCreativitySupport2014a} to evaluate how the image generators supported the participants' creative processes. Based on the recommendations by Cherry and Latulipe~\cite{cherryQuantifyingCreativitySupport2014a}, collaboration was marked as an optional item. This approach was adopted as participants were allowed to collaborate, but some opted not to do so actively.
After the ideation sessions, we conducted semi-structured group interviews focused on three main aspects: 1) how well the tools could produce the required images, 2) whether the tools provided novel solutions and 3) how participants thought the tools could be used in their design practice.
As we learned more about the tools in the first session, we also focused on participants' comments about an ``ideal'' tool that would support their design tasks. Two researchers conducted the interview, with one leading the discussion and another acting as a scribe, taking comprehensive notes.
The participant's comments during each session were written down by a researcher during the interview, and complemented with transcribed audio recordings. The full commentary consists of 2905 words and the audio recordings are approximately 50 minutes in total.
The commentary was analysed using content analysis~\cite{ContentAnalysis} to identify the participants' 
experiences using the tools.

\subsection{Image generation tools}%
The study was conducted with three text-to-image generation tools: Midjourney (version 4; \textbf{MJ}), DALL-E~2 (\textbf{DE}), and Stable Diffusion (version 1.5; \textbf{SD}).
These three tools represent the current state-of-the-art of text-to-image generation accessible to the public.
The tools have become very popular as they provide an easy-to-use means of synthesizing images from written text prompts.
    Midjourney and DALL-E are provided as web-based services by Midjourney and OpenAI, respectively.\footnote{https://www.midjourney.com, https://labs.openai.com}
    For Stable Diffusion, we used the web-based \textit{Dream Studio} interface,\footnote{https://beta.dreamstudio.ai} as provided by Stability AI.
The use of each of the three systems (MJ, DE, SD) in the sessions was balanced between participants, with up to two participants individually using one of the tools in each session. Each account at the three image generation services was filled with sufficient credits to allow image generation during the session. Participants were allowed to use their own laptops, apart from participants assigned to use SD. Data from two SD participants (P3, P4) was lost during S1 when it was noticed that SD only stores prompt history in the participant's local browser history. Participants assigned to use SD in S2-S3 worked with a provided laptop that allows us to access the locally stored browser history. The data from S2-S3 is limited to only 100 of the last prompts as the SD does not store a complete history of prompts.

\subsection{Participants}

\begin{table}[!tb]
    \small\centering
    \caption{The list of 17 study participants and their ages, genders, self reported years of studying architecture, study session, image generation tool, and prior experience with image generation.}
    \label{tab:participants}
    \begin{tabularx}{\textwidth}{XlllXrcX}
        \toprule
        No. & Age & Gender & Years studied & Session & Tool & Experience \\
        \bottomrule
        P1  & 22 & Male   & Three & S1 & DALL-E           & Yes \\ 
        P2  & 21 & Female & Three & S1 & DALL-E           & No \\
        P3  & 21 & Female & Three & S1 & Stable Diffusion & Yes \\
        P4  & 29 & Female & Two & S1 & Stable Diffusion & No \\
        P5  & 21 & Female & Three & S1 & Midjourney       & No \\
        P6  & 26 & Female & Five & S2 & DALL-E           & No \\
        P7  & 47 & Male   & More than five & S2 & DALL-E           & Yes \\
        P8  & 24 & Female & Three & S2 & Stable Diffusion & Yes \\
        P9  & 29 & Female & Seven & S2 & Stable Diffusion & No \\
        P10 & 24 & Male   & Four & S2 & Midjourney       & Yes \\
        P11 & 43 & Male   & One  & S2 & Midjourney       & Yes \\
        P12 & 25 & Female & Five  & S3 & DALL-E           & No \\
        P13 & 25 & Male   & Five & S3 & DALL-E           & Yes \\
        P14 & 23 & Female & Four & S3 & Stable Diffusion & No \\
        P15 & 24 & Male   & Five & S3 & Stable Diffusion & Yes \\ 
        P16 & 23 & Female & One & S3 & Midjourney       & No \\
        P17 & 24 & Female & Three & S3 & Midjourney       & No \\
        \midrule
        & M = 27 & 35.3\% Male & & & & 52\% No \\
        & StDev = 7 & 64.7\% Female & & & & 48\% Yes \\
        \bottomrule
    \end{tabularx}
\end{table}

We recruited 17 participants (P1--P17; 11 women, 6 men; self-reported genders) of ages 21--47 ($M=27$ years, $StDev=7$ years) using a university mailing list and instant messaging channels for architecture students. All participants were first to seventh-year architecture students, with a majority (35\%) being third-year students. 
Nine out of the 17 participants (52\%) had not used image generators previously, and only three (18\%) participants reported having used the systems five or more times.
Seven participants with prior experience stated they had used the tools just for fun and testing. Only one person said they had used it for visualization tasks. See \autoref{tab:participants} for the full description of the participants.

\section{Results}

In the three sessions, participants produced images with diverse prompts.
In the following,
we describe the generated images, analyse the prompt language used by the participants, and then the interview data, and general feedback during the sessions. In the qualitative section, we evaluate the efficacy of the image generators in facilitating the design task, examine the participants' utilization of prompts to visualize their ideas,
and discuss the qualitative insights gleaned from the group interviews.

\begin{figure}[!ht]
     \centering
     \begin{subfigure}[b]{\imageWidthSmall\textwidth}
         \centering
         \includegraphics[width=\textwidth]{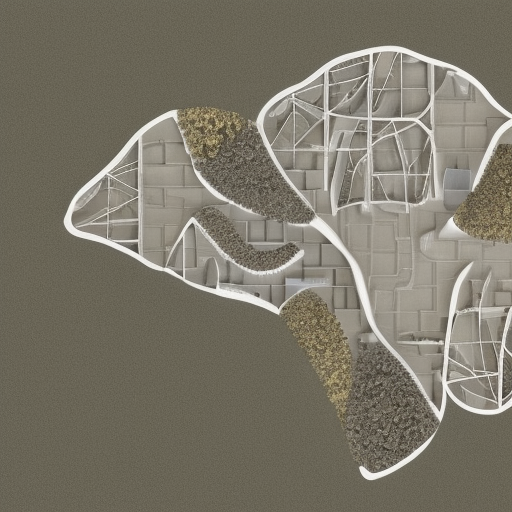}
         \caption{P4 floorplan}
         \label{fig:winning_images:a}
     \end{subfigure}
     \begin{subfigure}[b]{\imageWidthSmall\textwidth}
         \centering
         \includegraphics[width=\textwidth]{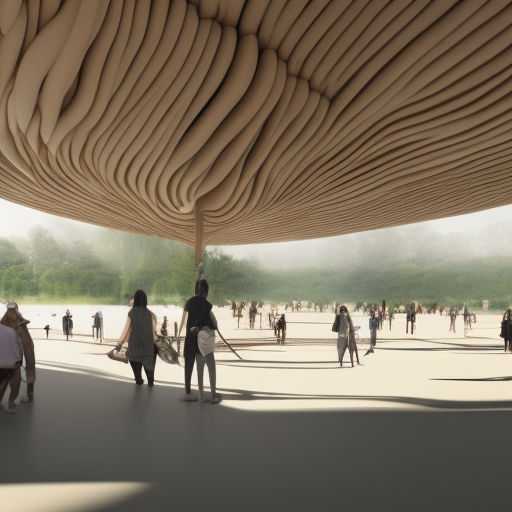}
         \caption{P4 interior}
         \label{fig:winning_images:b}
     \end{subfigure}
     \begin{subfigure}[b]{\imageWidthSmall\textwidth}
         \centering
         \includegraphics[width=\textwidth]{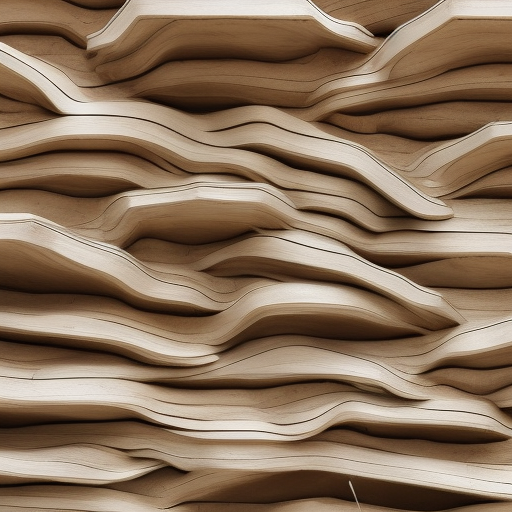}
         \caption{P4 material}
         \label{fig:winning_images:c}
     \end{subfigure}
    \\[6pt]     
     \begin{subfigure}[b]{\imageWidthSmall\textwidth}
         \centering
         \includegraphics[width=\textwidth]{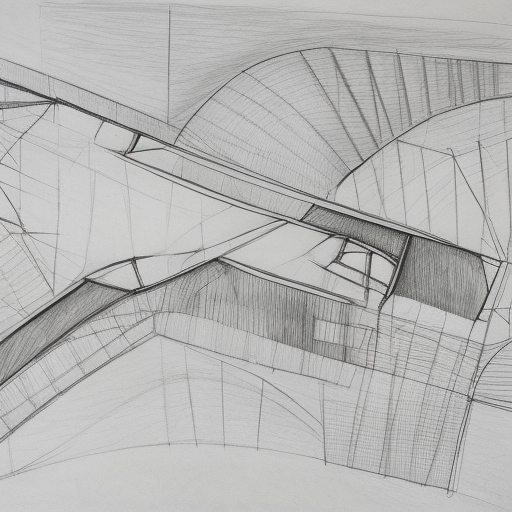}
         \caption{P9 floorplan}
         \label{fig:winning_images:d}
     \end{subfigure}
     \begin{subfigure}[b]{\imageWidthSmall\textwidth}
         \centering
         \includegraphics[width=\textwidth]{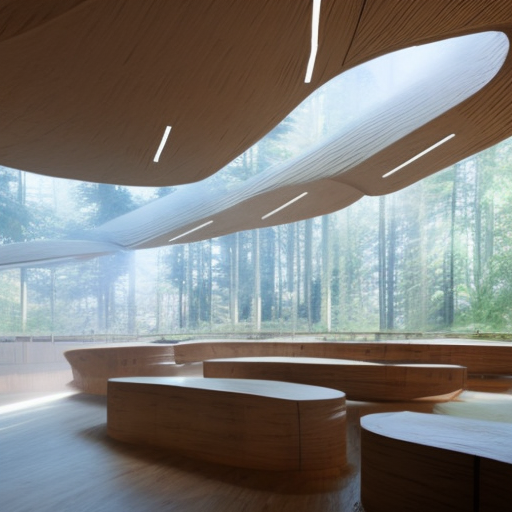}
         \caption{P9 interior}
         \label{fig:winning_images:e}
     \end{subfigure}
     \begin{subfigure}[b]{\imageWidthSmall\textwidth}
         \centering
         \includegraphics[width=\textwidth]{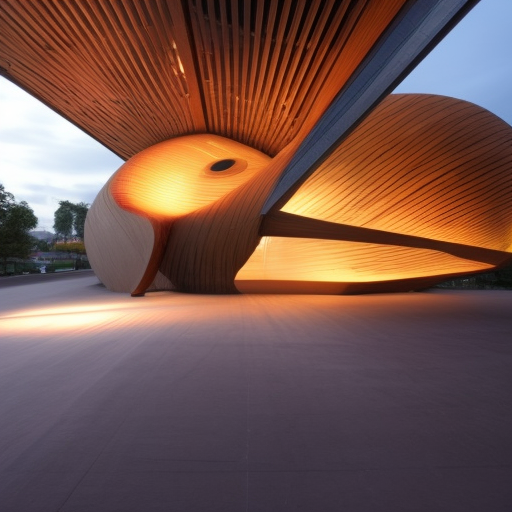}
         \caption{P9 material}
         \label{fig:winning_images:f}
     \end{subfigure}
    \\[6pt]     
     \begin{subfigure}[b]{\imageWidthSmall\textwidth}
         \centering
         \includegraphics[width=\textwidth]{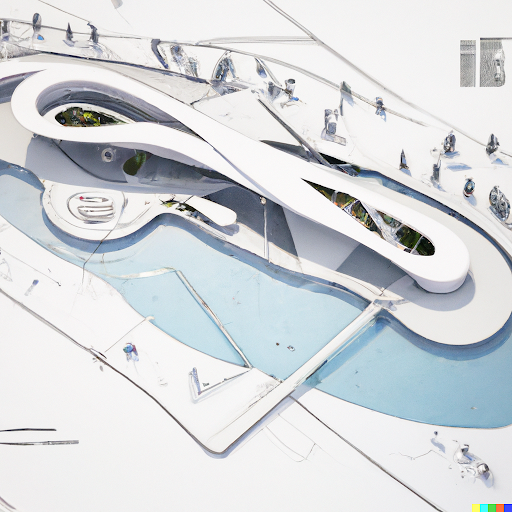}
         \caption{P12 floorplan}
         \label{fig:winning_images:g}
     \end{subfigure}
     \begin{subfigure}[b]{\imageWidthSmall\textwidth}
         \centering
         \includegraphics[width=\textwidth]{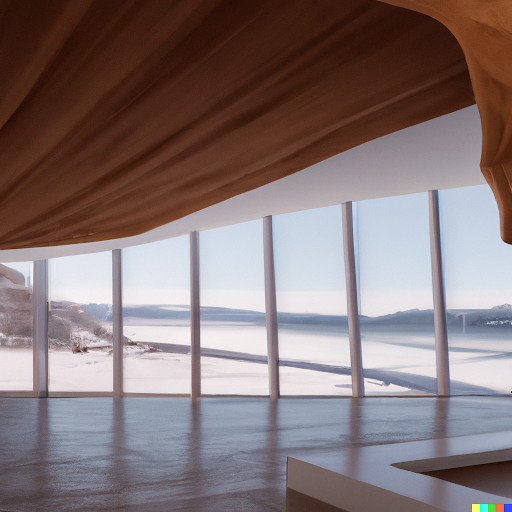}
         \caption{P12 interior}
         \label{fig:winning_images:h}
     \end{subfigure}
     \begin{subfigure}[b]{\imageWidthSmall\textwidth}
         \centering
         \includegraphics[width=\textwidth]{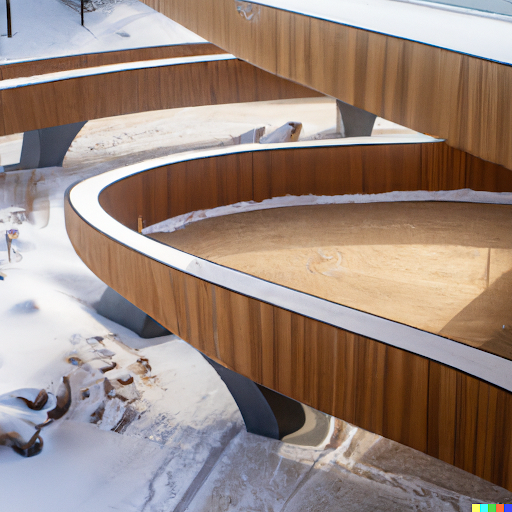}
         \caption{P12 material}
         \label{fig:winning_images:i}
     \end{subfigure}
    \caption{The three participants' floorplans, interior views, and facade materials voted best works from their respective sessions S1--S3. }
    \label{fig:winning_images}
\end{figure}

\subsection{Generated images}

Overall, the participants' concepts were understandable, as expressed through the generated images of floorplans, material samples, and indoor views. All participants were able to deliver all required images for their short presentation and had ample time for ideation and exploration during the session.

The participant-chosen winners of each session are depicted in \autoref{fig:winning_images}. Participants P4 (SD), P9 (SD), and P12 (DE) won in their respective sessions S1-S3. The winners' generated images suggest a strong sense of presence and shape language. Notably, all three winners focused on using wood in their designs.

Additionally, the winning designs use organic forms that would be difficult and time-consuming to model and produce with 3D modeling software. However, while the winning concepts use organic forms, that was not the case for all participants, as approximately half of the conceptual images used rectilinear forms. Then, some approached the design task through biophilic design language, taking inspiration from plants, trees, and mushrooms. For instance, the concept by P4 (see \autoref{fig:winning_images:d}) uses the scaled-up mushroom to convey space and atmosphere as if a person was standing under a mushroom. 

\begin{figure}[!ht]
     \centering
     \begin{subfigure}[b]{\imageWidthSmall\textwidth}
         \centering
         \includegraphics[width=\textwidth]{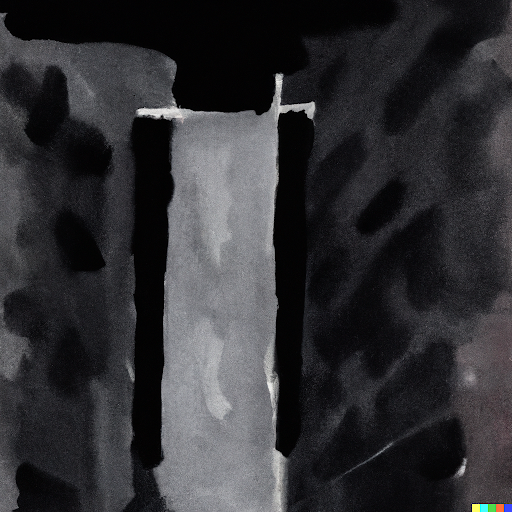}
         \caption{P1 floorplan}
         \label{fig:other_images:a}
     \end{subfigure}
     \hfill
     \begin{subfigure}[b]{\imageWidthSmall\textwidth}
         \centering
         \includegraphics[width=\textwidth]{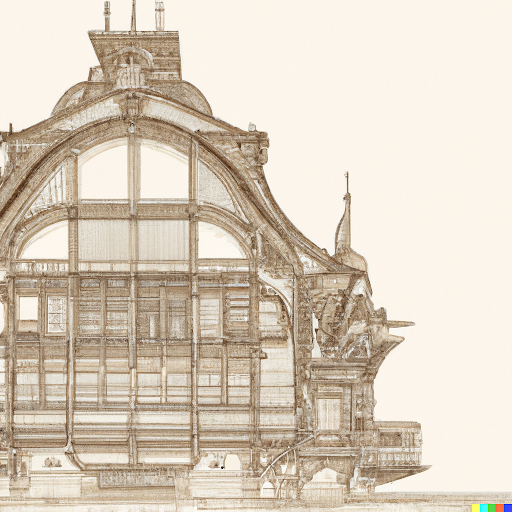}
         \caption{P7 material}
         \label{fig:other_images:b}
     \end{subfigure}
     \hfill
     \begin{subfigure}[b]{\imageWidthSmall\textwidth}
         \centering
         \includegraphics[width=\textwidth]{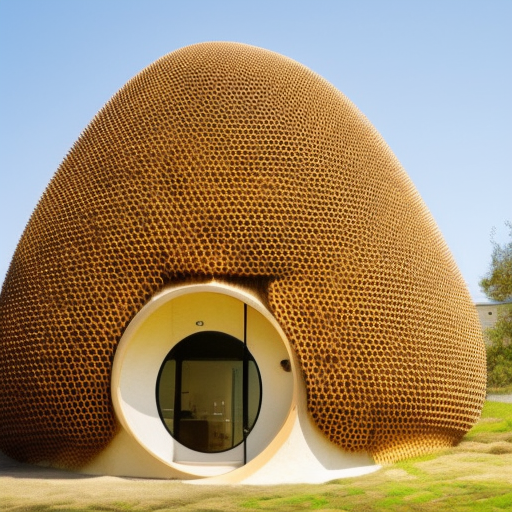}
         \caption{P9 oddity}
         \label{fig:other_images:c}
     \end{subfigure}
    \\[6pt]     
    \begin{subfigure}[b]{\imageWidthSmall\textwidth}
         \centering
         \includegraphics[width=\textwidth]{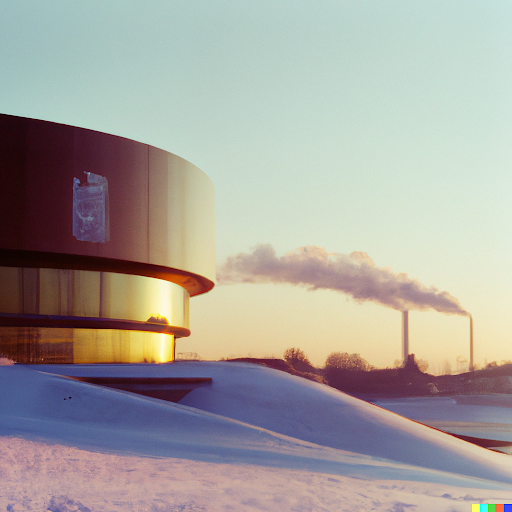}
         \caption{P13 material}
         \label{fig:other_images:d}
     \end{subfigure}
     \hfill
     \begin{subfigure}[b]{\imageWidthSmall\textwidth}
         \centering
         \includegraphics[width=\textwidth]{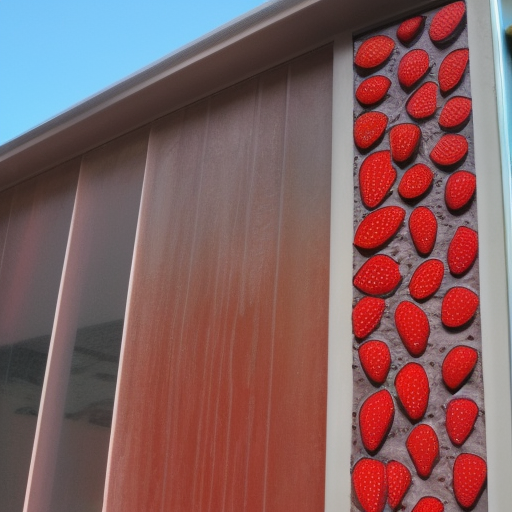}
         \caption{P3 material}
         \label{fig:other_images:e}
     \end{subfigure}
     \hfill
     \begin{subfigure}[b]{\imageWidthSmall\textwidth}
         \centering
         \includegraphics[width=\textwidth]{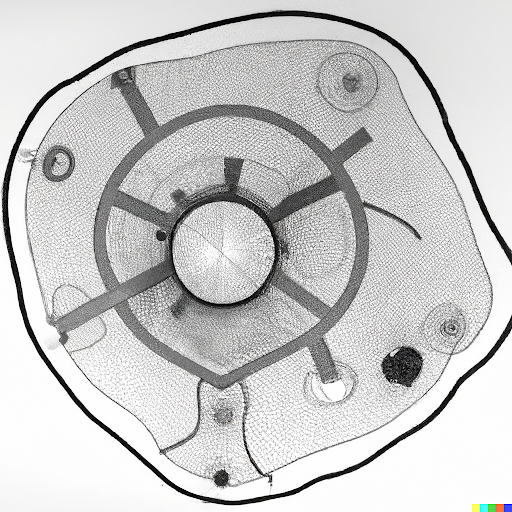}
         \caption{P13 floorplan}
         \label{fig:other_images:f}
     \end{subfigure}
    \caption{%
        Selection of images highlighting how the participants used image generators in unexpected ways. 
        In (a), P1 focused on representing a floorplan in an abstract style with a prompt ``watercolour plan view thick black walls.'' 
        In (b), P7 used a more ornate style for the facade. 
        In (c), P9 wanted to create a "honeycomb"-style material, which produced an actual honeycomb.
        In (d), P13 employed the design task's site by including factory chimneys, effectively suggesting location.
        In (e), P3 experimented with using strawberry as a facade material, and 
        in (f), P13 could not generate usable floorplans so they went for a more experimental approach.
    }
    \label{fig:other_images}
\end{figure}

Besides the winners, we highlight selected images that exemplify the participants' more adventurous experiences with the text-to-image generation tools in \autoref{fig:other_images}. Participants typically followed a straightforward methodology for generating their ideas, but especially at the beginning of each session, some exploration was conducted. For instance, P1 (see \autoref{fig:other_images:a}) prompted \textit{``watercolour plan view thick black walls''} to suggest a linear space with a sense of weight. P9 explored using abnormal construction methods, and the beehive-inspired building with prompt \textit{``spaceship-like''} (see \autoref{fig:other_images:c}) was as explorative as the spaceship-like buildings they explored with. 
In contrast to the organic approaches, P7 opted for a more ornate style (see \autoref{fig:other_images:b}), implying a more traditional approach with decorative elements. However, to what extent these decorations are meaningful in the designs is an open question.

Generating floorplans proved to be especially challenging (see the floorplans in \autoref{fig:winning_images} and \autoref{fig:other_images}). The floorplans were rarely the black-and-white plan drawings that participants had learned to expect and instead were usually colored, three-dimensional, or had nonsensible layouts (e.g. rooms with sharp corners or missing doors). 
Material samples were also challenging, as the image generators were ineffective in generating facades. Thus, many participants defaulted to simple images of the material or used exterior perspective views.
At times the participants had to settle for images that did not meet their criteria, which can often occur when building prompts~\cite{oppenlaender2023prompting}.

\begin{table}[bt]
    \small\centering
    \caption{The mean and standard deviation values for the counts, scores, and weighted scores for each of the six subfactors in the CSI, across all three image generators.
    The mean factor counts are the number of times that participants chose that particular factor as important to the task (out of six factors). The factor scores range between 2 (worse) and 20 (better).
    Weighted factor scores are calculated by multiplying a participant’s factor agreement scale score by the factor count.}
    \begin{tabularx}{\textwidth}{lrrr}
        \toprule
        Scale & Mean Factor Counts (StDev) & Mean Factor Score (StDev) & Mean Weighted Factor Score (StDev) \\
        \midrule
         Collaboration      & 0.47 (0.87)           & 6.88 (5.24)           & 3.24 (4.58)  \\ 
         Enjoyment          & 2.53 (1.18)           & 17.1 (1.96)           & 43.30 (2.32) \\ 
         Exploration        & 4.41 (0.62)           & 14.1 (2.87)           & 62.28 (1.77) \\ 
         Expressiveness     & 3.00 (1.27)           & 14.9 (3.21)           & 44.82 (4.09) \\ 
         Immersion          & 2.12 (1.17)           & 11.5 (3.89)           & 24.29 (4.54) \\ 
         ResultsWorthEffort & 2.47 (1.55)           & 15.7 (2.42)           & 38.80 (3.74) \\ 
         \bottomrule
    \end{tabularx}
    \label{tab:csi_factors}
\end{table}

\subsection{Creativity support}%
The CSI scores were calculated with the method described by Cherry and Latulipe~\cite{cherryQuantifyingCreativitySupport2014a} by summing the weighted subfactors. A Kruskal-Wallis test found no significant differences in the CSI scores between the three generative tools ($\chi^{2}(2)=1.04, p=0.59$): DE ($M=75.2$, $StDev=13.4$);~MJ ($M=70.9$, $StDev=15.8$); and SD ($M=73.7$, $StDev=8.9$). Additionally, a one-way ANOVA showed that the six subfactors had significant differences in the count means ($F(5)=21.07, p<0.001$), score means ($F(5)=19.64, p<0.001$), and weighted mean scores ($F(5)=18.35, p<0.001$).
Analyzing the subfactors in \autoref{tab:csi_factors}, we see that participants most valued ``Exploration'' and found the generative tools most supported ``Enjoyment.'' Eight participants found the ``Collaboration'' subfactor inapplicable, resulting in a lower score for that factor. 

\subsection{Prompts}%
Participants wrote a total of 588 prompts during the three sessions with an average of 39.2 prompts per participant ($StDev=19.6$).
The prompts contained between 2 and 53 tokens (discrete words), with an average of 15.4 tokens per prompt ($StDev=6.9$ tokens).


%
The participants' prompts were overall descriptive, using appropriate terms to describe an intended outcome. Some examples of prompts include:%
\begin{itemize}
    \item architectural floorplan of concert hall and a grand staircase (P1,~DE)
    \item elevation detail photo of brutalist style culture center in winter time (P11,~MJ)
    \item floor plan of zaha hadid wooden culture center over the estuary in bright winter day (P12,~DE)
    \item akosnometric explosion image of cultural centre floorplans (P13,~DE)
    \item architecture, floorplan, round, realistic materials, cultural center, HQ, 4k (P14, SD)
    \item experiential architectural wooden pavilion building in the park (P17,~MJ)
\end{itemize}

\subsubsection{Prompt sequences}
\begin{figure}[!htb]
\centering
 \includegraphics[width=0.88\textwidth]{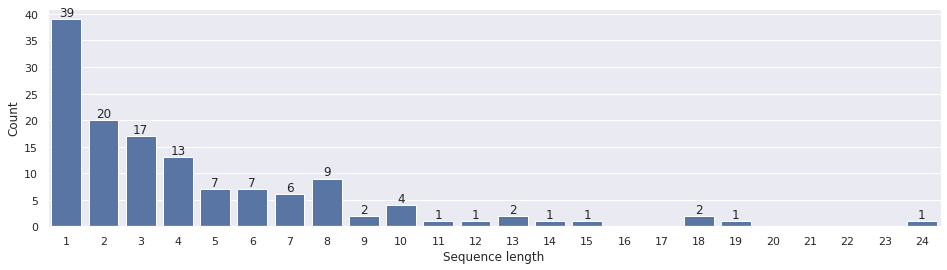}
\caption{The length of participants' prompt sequences demonstrates the commitment of participants to a train of thought during the ideation session. Most ideas would spawn at least a few ($<$4) prompts before the participant moved to a new prompt sequence.}
\label{fig:sequences}
\end{figure}

Participants would typically only start over with completely new prompts when results were unsatisfactory or when starting another task. We manually grouped similar prompts together, and identified breaks where participant switched from one sequence of prompts to another. These sequences (i.e., groupings of prompts belonging to one idea) consist of an initial prompt that was iteratively extended by the participant with keywords to improve the resulting images (see \autoref{fig:sequences}). On occasion, participants would return to previous sequences later on, e.g., Seq2 $\rightarrow{}$ Seq3 $\rightarrow{}$ Seq4 $\rightarrow{}$ Seq2.
The following is an excerpt of a prompt sequence P13-Seq1 (DE) with changes in between prompts highlighted:%
\begin{quote}
1. modernistic and ecological cultural centre in a waterfront environment with people standing in the foreground of the image\\
2. modernistic\hl{\textit{, abstract}} and ecological cultural centre in a waterfront environment with people standing in the foreground of the image \hl{\textit{in a sunset}}\\
3. modernistic, abstract and ecological cultural centre in a waterfront environment \hl{\textit{in a island}} with people standing in the foreground of the image in a sunset\\
4. modernistic, abstract and ecological cultural centre in a waterfront environment in a island with people standing in the foreground of the image in a sunset \hl{\textit{digital art}}
\end{quote}
The sequences of prompts were up to 24~prompts in length (for P10,~MJ) with an average sequence length of 8.4 prompts ($StDev=5.9$ prompts).
The average length of the sequences (see \autoref{fig:sequences}) demonstrates the participants' engagement with the image generation tool to solve the given design task.



\subsubsection{Prompt language}
\begin{figure}[!htb]
\centering
 \includegraphics[width=\textwidth]{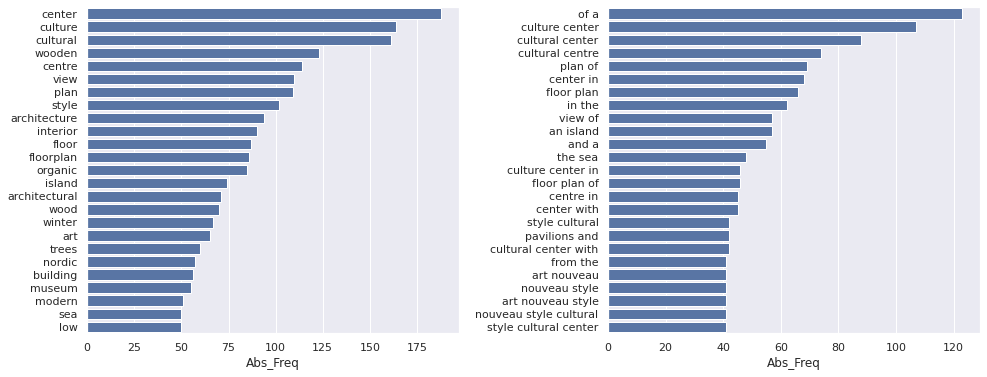}
\caption{Most frequently used tokens in participant-written prompts. The plot on the left depicts the 25 most frequent tokens with stop words removed. The plot on the right depicts the 25 most frequent n-grams.}
\label{fig:tokens}
\end{figure}

\autoref{fig:tokens} depicts the most common keywords and keyword combinations used by participants.
    The keywords largely focus on the design task given to participants: creating a concept for a regional culture center.
    We find that most participants adopted the language used in the design brief, as evident in their prompts. For instance, `floorplan' and `facade' were often used. Only few participants veered off from this given path and experimented with other terms, such as synonyms (e.g., using `wireframe' or `layout' instead of floorplan).
As for the use of `prompt modifiers' commonly used by practitioners of text-to-image generation~\cite{TTI-taxonomy,TTI-creativity},
a minority of participants used such modifiers to enhance their images.
These modifiers include, for instance, HQ, 4k, photo realistic (P14, SD) and 8k, unreal engine (P15, SD).
Names of architects were used very sparingly by participants in their prompts.
The three names of architects were
\textit{``frank lloyd wright''} (used by P12 (DE) in two prompts), \textit{``zaha hadid''} (by P12 (DE) in 18 prompts), and
\textit{``renzo piano''} (by P6 (DE) in 33 prompts).

\subsection{Qualitative insights}

During the introduction, the participants quickly learned how to build basic prompts, and when working on the design task, the participants learned strategies that supported their creative goals. Participants stated the image generators were ``\textit{easy and fun to use}'' (P6,~DE), it was ``\textit{really wonderful to be inspired and get ideas}'' (P5,~MJ), and they were ``\textit{a bit like an extension of your own imagination}'' (P2,~DE). For some, the tools were surprisingly useful, e.g. ``\textit{\ldots this worked better than I thought, I didn’t expect how creative and great things can come from this}'' (P4,~SD). Some participants found themselves in a flow state when working with the tools: ``\textit{you get lost in there very easily, and forget what you were supposed to do, you just ideate and generate more and more}'' (P5,~MJ), which was also evident in the CSI scores as ``Exploration'' was the most important subfactor. 
Analyzing the interview data, we constructed topics describing how the participants considered images in the design process, randomness, strategy, and recommendations for image generators.

\subsubsection{Synthetic images in the architectural ideation process}
The participants in all three groups had different views on how the generated images supported their architectural design and ideation processes. 
As the participants had different backgrounds and amounts of experience with architectural design, their needs for creativity support tools varied. 
To understand how image generators can support their creative process, the participants compared image generation to familiar applications. The most frequent comparison was to Pinterest, a website that the students use as a source of inspiration for architectural ideation. One benefit of image generation was the capability to produce unique images that are closer to what the person aims for with their specific design concept -- \textit{``you can get your own idea so much more precisely here than if you used some existing images''} (P3,~SD).
%

While the tools were useful for generating images, participants had mixed opinions on how the tools and the images could be used for ideation. One participant explained that relying on image generation alone is not meaningful for ideation, ``\textit{because architecture is much more than what something looks like}'' (P17,~MJ). Beyond inspiration from images, the design process is also concerned with meeting the needs of different stakeholders, and for some participants, image generation does not provide a suitable foundation for design.

However, participants had mixed opinions about how generative software can be part of a design process. On the one hand, the designer needs to actively make sense of the systems and how they are useful for meeting design goals. On the other hand, the systems were recognized to have an influence on the design process. As such, while the image generators provided flow-state creative experiences, image generators were not seen as neutral or value-free tools. As architectural design processes have grown ever more digital, in the future, architects require a better understanding of the value of the image generators:


\begin{quote}%
    \textit{When you have an architectural background, you clearly notice that what is really needed is a person's thoughts for requesting things; the machine cannot do what the user wants without guidance.} (P7,~DE)
\end{quote}

\subsubsection{Randomness and sensemaking}




The participants found the tools produced results that seemed random to what they expected from the prompts. Using image generation was seen as ``\textit{absurd}'' (P2,~DE) and ``\textit{abstract}'' (P7,~DE), and the participants did not always recognize how they could control the randomness. As such, participants had mixed thoughts about how -- or whether -- to control randomness in future generative design tools. Controlling randomness means more predictable results and enables focusing more precisely on one's design vision. Randomness was also used constructively to support one's creative process with greater diversity. One participant stated:



\begin{quote}
    \textit{It's nice when artificial intelligence cares little about what you wrote, and the same can be applied the other way around, that you can care little about what it's trying to generate and you try to come up with some alternative explanations for them other than what would be the first thought. \ldots Now I feel like I should have looked for things less and let the system produce more, and just tried how it could go wrong. And you can figure it out yourself if it makes any sense because the system doesn’t put the sense in there.} (P1,~DE)
\end{quote}




\subsubsection{Learning strategies}

In order to generate meaningful images, participants implemented different strategies. Using relevant vocabulary was a key aspect. Participants found themselves using words and phrases which they use in their usual design processes but were not conducive to generating useful images. As floorplans were found especially tricky to generate, participants attempted to get better results or interesting variations by using alternative words such as \textit{``plan drawing''} and \textit{``plan view,''} or by changing the word order.
In this sense, using image generation required the participants to consider the system's inner workings and use alternative keywords: ``\textit{You would have to be more creative and forget the architectural terms}'' (P4,~SD).
The participants' struggles in identifying robust sets of keywords
suggest that the participants
tried to align their ideation processes with
what they expected to be suitable inputs to the image generation tool,
instead of relying on their own creativity and learned design practice.

The length of the prompts was another point of strategizing. Participants had differing opinions on how the applications handled the prompts. For instance, P5~(MJ) found that modifiers like \textit{``lots of''} were not that successful but also found that the first words in the prompt were more impactful than those in the end. Additionally, P3~(SD) learned that longer sets of keywords were more successful than shorter ones, although longer prompts sometimes did not take all of the keywords into account.
Finally, to generate suitable images, many participants also worked iteratively by adding keywords to refine their designs. For instance, P6's statement suggests they understood that one could learn to construct better prompts and control image generation with more expertise:%
\begin{quote}%
    \textit{I tried to generate images that I was already quite satisfied with, and then I mostly changed smaller things, such as the shape of the roof. But I think if you had used the application more, you would know how to change things more skillfully.} (P6,~DE)
\end{quote}%


\subsubsection{Participants' improvement recommendations}

Drawing inspiration from other applications, such as Pinterest, one participant found value in being able to have a curated algorithm that supports a person's personal style. Liking certain generated images could help this algorithm to produce similar images or recommend other generated images.
Additionally, refining some specific parts of the image was seen as useful. While advanced editing features of the image generators, such as inpainting and outpainting, were not used in our study, it is interesting that participants recognized the utility of such features. Being able to specify and re-generate certain parts of the image would help to form a more precise concept.


Finally, some participants valued having constraints in the system. Architecture often works around the many parameters brought by the site's context, so having them present would help the creative ideation process. For instance, one participant suggested: ``\textit{You could put some general size, some number of square meters, number of floors, details or more specific specs and it will generate suitable things}'' (P6,~DE). Statements like these show the utility of embedding traditional computer-aided design software with generative design features.

\subsection{Unexpected results and prompting challenges}

The participants faced several struggles while building their prompts, some of which were rather unexpected. Many participants tried to visualize ideas using references to local buildings and city features. These features are unlikely to be recognized by the image generation system, and while the generated images included parts of such buildings (e.g., factory pipes), features (e.g., islands or bridges), or landmarks, the images would not reflect the actual location of the design task in any meaningful way.
For instance, P5 (MJ) tried the prompt of a \textit{``culture center [...] in [ANONAREA] [ANONCITY] [ANONCOUNTRY]''}
but realized the image generation system would not generate identifiable local landmarks and subsequently abandoned the prompt. 

Another struggle of participants was removing objects and text from the generated images. Negatively weighted prompts~\citep{TTI-taxonomy} are a technical feature of image generation systems that can address this shortcoming.
However, very few participants made use of this feature, and others tried to emulate the feature in their prompts, adding terms such as \textit{``without text''} (P1,~DE) or \textit{``not norwegian style''} (P11,~MJ).
Additionally, architects often seek inspiration from nature, and many participants wrote prompts inspired by natural organic structures and materials. Many used quite literal language in the prompts, and sometimes the addition of such inspirational terms led to unexpected results.
For instance, P9 (SD) experimented with beehive-like buildings and ended up with an unusual structure (see \autoref{fig:other_images:c}). While interesting, these material choices could be difficult to manufacture, and odd-looking structures could either not be structurally sound or be otherwise unfeasible for the intended purpose.

While participants typically pursued a coherent idea, but occasionally after exploring many variations of the same prompt, participants noticed they were going down a path that did not lead to the expected results. These participants regressed to earlier versions of their own prompts. 
This implies that even with fine-tuning prompts, it is unlikely that the user can achieve exactly what they want. 
While the image quality can be improved with skillful text prompting, it can also be a limiting factor in the creative process if overused.

\section{Discussion}

One implication of our study is that text-to-image generators could offer ideation tools in the context of creativity in architecture.
In current systems, a lot of effort has to be put into the prompt language, as the choice of words matters a lot~\cite{2303.04587.pdf}.
Recent research in the field of Human-Computer Interaction (HCI) also points out text prompting as a skill with a learning curve~\cite{oppenlaender2023prompting}.
Typically longer prompts lead to higher quality results~\cite{2303.04587.pdf}.
Chang et al. describe prompt writing as an art form and that "artists" - in our case, designers - also benefit from being highly skilled with natural language~\cite{2303.12253.pdf}. And indeed, when using AI for design, ``A Word is Worth a Thousand Pictures,'' as Kulkarni et al. eloquently phrased it in the title of their work~\cite{2303.12647.pdf}.

\subsection{Ideation and creativity with AI}

In a 2013 lecture, the renowned contemporary architect Bernard Tschumi explained the role of visual media in architecture as ``\textit{It is not about the images, it is what the images do}.''~\cite{tschumi2013Red}. As architecture can be seen as a field that communicates through images, it is the effect of the images that matter and what political and social impact they have. Therefore, image generation could be understood as a part of a more context-aware creative practice. For instance, theories on situated cognition propose that human knowledge is bound to the physical and social context, which are also applicable for creativity~\cite{malininCreativePracticesEmbodied2016}.
 Crucially, this understanding of creativity places importance on \textit{where} creativity happens and shows how creativity is contextual. As architecture education is suggested to focus more on students' reflective skills for creativity~\cite{parkCreativeThinkingArchitecture2022}, we posit that the usage of image generators requires that the students engage them with situated cognition and creativity in mind. In other words, this requires the students to understand their own motivations, skills, or abilities as well as the affordances of the chosen tools. Returning back to Tschumi's quote, we can ask in what ways the images produced by AI can facilitate the designer's reflective skills for creativity or provide new affordances for exploration and seeing the design task in a new way. Additionally, in what ways does image generation respond to prompting that stems from certain architectural traditions? In order to utilize image generation effectively in the situated model of creativity~\cite{tanggaardSituatedModelCreative2014}, the generative systems need to be more explicit about the training data sets and their inherent biases.

\subsection{Use of AI in the architectural design process - lessons learned}
Through our design tasks with image generators, we learned how the students adopted the image generators and how our findings can inform the use of generative methods in the future. To this end, we propose points of consideration for developing image generators for the domain of architecture and practices for adopting image generation in architectural education.

\subsubsection{Considerations for the design of image generators.}
The image generators we used supported the participants' creativity equally. However, for architectural design purposes, we point out two main aspects that can improve the experience of concept ideation.

Our study shows that the image generators failed to generate conventional floorplan drawings. We suggest that due to the fact that floorplans are more abstract representations than e.g., perspective visualizations, the training dataset of floorplans images has insufficient labeling for architectural design purposes. While generative methods have been popular~\cite{buhamdanGenerativeSystemsArchitecture2021}, fine-tuned language models can support generating more refined floorplans through text~\cite{galanos2023architext}. However, the availability of architectural floor plan datasets is still an open issue, as well as the appropriate size of training data and data curation for ``\textit{features like cultural habits, economy, climate conditions, constructive systems, legislative criteria}.''~\cite{rodriguesGeneratingFloorPlans2022}

Additionally, the participants found that it was difficult to track the flow of different ideas in the vertically scrolling interfaces of the image generators. Participants suggested that using a more spatial layout would help to explore different aspects of the design concept while also being able to zoom out and see how the ideation has progressed.

\subsubsection{Considerations for educators}
Many researchers believe that large language models (LLMs), such as ChatGPT and GPT-4, will affect how education is carried out in Higher Education. LLMs can be applied for general problem solving and can act as personal intelligent tutors~\cite{ethanmollick-isthereabetterquote}.
This new capacity for solving given design tasks could also affect how architecture is taught in Higher Education. The capacity of LLMs to respond to natural language prompts is a result of their emergent behavior~\cite{2206.07682.pdf}. Prompt-based interaction with the LLM is a means to trigger this emergent behavior.
Hovever, careful consideration of the value of image generation is needed. Through observations on student design skills after adopting CAD techniques, Meneely and Danko suggested that design education should focus on the question of `why' to promote reflection on the usefulness of different technologies for the design process~\cite{meneelyMotiveMindMedia2007}. The authors state: ``\textit{If we cannot articulate or justify what leads us to technology in the first place, then how do we determine if technology is truly serving us?}''~\cite{meneelyMotiveMindMedia2007}.
Therefore, using image generation in education should start with the motivation for using them, and after that, teaching the effective use with the right vocabulary and for appropriate design tasks.
Architecture education should focus on teaching students a detailed and broadly applicable expert vocabulary needed to describe envisioned design concepts to generative models effectively. 
For instance, our study shows that while students are able to describe design concepts in descriptive language, many participants did not apply prompt modifiers (cf. \autoref{fig:tokens}).
Educating the limitations of state-of-the-art methods, highlighting the trade-off, and explaining when it is a good idea to use them and when more traditional methods are preferred helps to build generative architectural design expertise.

\subsection{Limitations}
We acknowledge that text-to-image generators can be used very effectively, especially by expert users and when using advanced features. In our study setting, the participants were largely inexperienced and started using them from scratch, at least in the context of the design task. Better instructions or extended tasks could improve the generated images and help sort out the challenges that came up in our experiment.

\section{Conclusions}
The recent rapid development in image generation has the potential to transform the design processes in architecture -- a field heavily concerned with the production of visual media. 
We conducted a laboratory study with 17 architecture students to understand how they adopted image generation in the early stages of architectural concept ideation. 
Using standardized questionnaires on creativity support, prompt analysis, and group interviews, we learned that the participants approached image generators with different creative mindsets. 
In order for image generators to be effective and meaningful in architectural design, the design of the image generators needs to support creative exploration, and architectural educators need to emphasize appropriate usage and teach advanced usage. 

\end{document}